\documentclass{article}
\usepackage{graphicx}
\usepackage{amsxtra,amssymb,amsthm,amsmath,latexsym}

\title{Modeling micro-macro pedestrian counterflow in heterogeneous domains}
\author{Joep Evers\thanks{Corresponding author.} \thanks{Department of Mathematics and Computer Science, PO Box 513, 5600 MB Eindhoven, TU Eindhoven, The Netherlands. E-mail:  \texttt{j.h.m.evers@student.tue.nl}}
\and Adrian Muntean\thanks{CASA -- Centre for Analysis, Scientific computing and Applications, Department of Mathematics and Computer Science, Institute for Complex Molecular Systems (ICMS), TU Eindhoven, PO Box 513, 5600 MB Eindhoven, The Netherlands. E-mail:  \texttt{a.muntean@tue.nl} }}

\newtheorem{postulate}{Postulate}[section]
\newtheorem{definition}{Definition}[section]
\begin{document}
\maketitle
\begin{abstract}
We present a  micro-macro strategy able to describe the dynamics of crowds
in heterogeneous media. Herein we focus on the example of pedestrian
counterflow. The main working tools include the use of mass and
porosity measures together with their transport as well as suitable
application of a version of Radon-Nikodym Theorem formulated for
finite measures. Finally, we illustrate numerically our microscopic model
and emphasize the effects produced by an implicitly defined {\em
social} velocity.

{\bf PACS 2010} : 89.75.Fb;
02.30.Cj;
02.60.Cb;
47.10.ab;
45.50.Jf;
47.56.+r

\end{abstract}


\section{Introduction}
One of the most annoying examples of collective behavior\footnote{See the question of scale of Vicsek \cite{Vicsek}.} is
pedestrian jams --  people get clogged up together and cannot reach
within the desired time the target destination. Such jams are the
immediate consequence of the {\em simple exclusion process}
\cite{Landim,Schadschneider}, which basically says that two
individuals cannot occupy the same position $x\in \Omega\subset
\mathbb{R}^d$ at the same time $t\in S:=]0,T[$, where $T\in
]0,\infty[$ is the final moment at which we are still observing our
social network.

Observational data (cf. e.g. \cite{Kretz}) clearly indicates that such
jams typically take place in certain neighborhoods of
bottlenecks\footnote{Bottlenecks are places where people have a
reduced capacity to accommodate locally \cite{Schadschneider}.}
(narrow corridors, exits, corners, inner obstacles/pillars, ...).
The effect of heterogeneities\footnote{Note that, for instance, Campanella et al. \cite{Campanella} give a different meaning to {\em heterogeneity}: they mainly refer to lack of homogeneity in the speed distributions of pedestrians. In \cite{Campanella0} the geometric heterogeneities - obstacles - are introduced in the microscopic model.} on the overall dynamics of the crowd
is what motivates our work.

\begin{figure}
\begin{center}
\includegraphics[width=8cm]{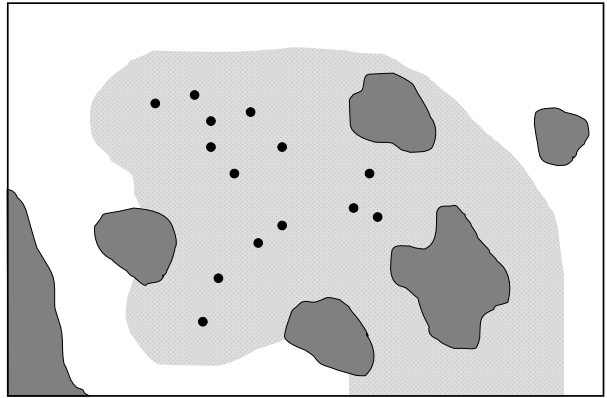}\caption{Schematic representation of the heterogeneous medium $\Omega$. The little black  discs represent the pedestrians, while the dark gray zones are the parts where the pedestrians cannot penetrate (i.e. subsets of $\Omega_s$). The pedestrians are considered here to be the microscopic entities, while the grayish shadow indicates a macroscopic crowd; see Section \ref{section Mass measure} for the precise distinction between {\em micro} and {\em macro} made in terms of supports of micro and macro measures.}
\label{geometry}
\end{center}
\end{figure}

In this paper we start off with the assumption that inside a given
room (e.g.  a shopping mall), which we denote by $\Omega$, there are
{\em a priori} known zones with restricted access for pedestrians (e.g.
closed rooms, prohibited access areas,  inner concrete
structures)\footnote{Note that some neighborhoods of these places can host, with a rather high probability, congestions!}, whose union we call $\Omega_s$.
Let us also assume that the remaining region, say $\Omega_p$, which is defined by
$\Omega_p:=\Omega-\Omega_s$, is connected. Consequently, $\Omega_p$
is accessible to pedestrians. The exits of $\Omega$ -- target that
each pedestrian wants to reach -- are assumed to belong to the boundary of
$\Omega_p$. The way we imagine the heterogeneity of $\Omega$
 is sketched in Figure \ref{geometry}.

In this framework, we choose for the following working plan:
Firstly, we extend the multiscale approach developed by Piccoli et
al. \cite{Piccoli2010} (see also the context described in
\cite{PiccoliTosin} and \cite{PiccoliTosinMeasTh}) to the case of
counterflow\footnote{Two groups of people are moving in opposite
directions.} of pedestrians; then we allow the pedestrian dynamics
to  take place in the heterogeneous domain $\Omega$, and finally, we
include an implicit velocity law for the pedestrians motion. The
main reason why we choose the counterflow scenario [also called
bidirectional flow \cite{Campanella0}] out of the many other
well-studied crowd dynamics scenarios  is at least threefold:
\begin{itemize}
\item[(i)] Pedestrians counterflow is often encountered in the
everyday life: at pedestrian traffic lights, or just observe next
week-end, when you go shopping, the dynamics of people coming against your walking direction [especially if you are positioned inside narrow
corridors].
\item[(ii)] The walkers trying to move faster by avoiding local interactions with the oncoming
pedestrians facilitate the occurrence of a well-known self-organized
macroscopic pattern -- lane formation; see, for instance \cite{HelbingVicsek}.
\item[(iii)] We expect the solution to microscopic models posed in narrow
corridors to be computationally cheap. Consequently, extensive
sensitivity analyses can be performed and the corresponding
simulation results can be in principle tested against existing
experimental observations \cite{Kretz,Campanella}.
\end{itemize}
The presence of heterogeneities is quite natural. Pedestrians
typically follow existing streets, walking paths, they trust
building maps, etc. They take into account the local environment of
the place where they are located. If the number of pedestrians is relatively high compared to the available walking space, then the  {\em crowd-structure interaction} becomes of vital importance; see e.g. \cite{Bruno} for preliminary results in this direction.

As long term plan, we wish to understand what are the microscopic
mechanisms behind the formation of lanes in heterogeneous
environments. In other words, we aim at identifying links between
{\em social force}-type microscopic models (see \cite{HelbingMolnar,Maury}, e.g.) and macroscopic models
for lanes (see \cite{HelbingVicsek,Bellomo}, e.g.) in the presence of heterogeneities. Here we follow a measure-theoretical approach to describe the dynamics of crowds\footnote{The pedestrians are not exposed here to panic situations.}. Our working strategy is very much inspired by  the works by M. B\"ohm \cite{Boehm} and  Piccoli et al. \cite{Piccoli2010}.

The paper is organized as follows: In Section \ref{measure} we
introduce basic modeling concepts defining the mass and porosity
measures needed here, as well as a coupled system of transport
equations for measures. In Section \ref{social} we present our
concept of {\em social velocity}. Section \ref{micro-macro}
contains the main result of our paper -- the weak formulation of a
micro-macro system for pedestrians moving in heterogeneous domains.
We close the paper with a numerical illustration of our
microscopic model (Section \ref{simulation}) exhibiting effects induced by an implicitly defined velocity.

\section{Modeling with mass measures. The porosity
measure}\label{measure}

  For basic concepts of measure theory and their interplay with modeling in materials and life science, we refer the reader, for instance, to \cite{Halmos} and respectively to \cite{Boehm,PiccoliTosin,interactions}.

\subsection{Mass measure}\label{section Mass measure}

Let $\Omega\subset\mathbb{R}^d$ be a domain (read: object, body)
with mass. Since we have in mind physically relevant situations
only, we consider $d\in\{1,2,3\}$. However, most of the
considerations reported here do not depend  on the choice of the
space dimension $d$. Let $\mu_m(\Omega')$ be defined as the mass in
$\Omega'\subset\Omega$. Note that whenever we write
$\Omega'\subset\Omega$, we actually mean that $\Omega'$ is such that
$\Omega'\in\mathcal{B}(\Omega)$, where $\mathcal{B}(\Omega)$ the $\sigma$-algebra of the Borel subsets of $\Omega$. As a rule, we assume $\mu_m$ to be
defined on all elements of $\mathcal{B}(\Omega)$.

In Sections \ref{section micro mass measure} and \ref{section macro
mass measure}, we consider two specific interpretations of this mass
measure that we need to describe  the behavior of pedestrians at two
separated spatial scales.

\subsubsection{Microscopic mass measure}\label{section micro mass measure}

Suppose that $\Omega$ contains a collection of $N$ point masses (each of them of mass scaled to 1), and denote their positions by $\{p_k\}_{k=1}^N\subset\Omega$, for $N\in\mathbb{N}$. We want $\mu_m$ to be a counting measure (see Sect. 1.2.4 in  \cite{Ash}, e.g.) with respect to these point masses, i.e. for all $\Omega'\in\mathcal{B}(\Omega)$:
\begin{equation}
\mu_m(\Omega')=\#\{p_k\in \Omega'\}.
\end{equation}
This can be achieved by representing $\mu_m$ as the sum of Dirac
measures, with their singularities located at the $p_k$, $k\in
\{1,2,\ldots,N\}$, namely:
\begin{equation}\label{mum}
\mu_m=\sum_{k=1}^{N} \delta_{p_k}.
\end{equation}
We refer to the measure $\mu_m$ defined by (\ref{mum}) as \textit{microscopic mass measure}.

\subsubsection{Macroscopic mass measure}\label{section macro mass measure}

Let us now consider another example of mass measure $\mu_m$. To do this, we assume that the following postulate applies to $\mu_m$:
\begin{postulate}[Assumptions on $\mu_m$]\label{Postulate mu_m}
\begin{enumerate}
  \item[(i)] $\mu_m\geqslant 0$.\label{Postulate mu_m 1}\\
  \item[(ii)] $\mu_m$ is $\sigma$-additive.\label{Postulate mu_m 2}\\
  \item[(iii)] $\mu_m\ll\lambda^d$, where $\lambda^d$ is the Lebesgue-measure in $\mathbb{R}^d$.\label{Postulate mu_m 3}
\end{enumerate}
\end{postulate}
By Postulate \ref{Postulate mu_m} (i) and (ii), we have that
$\mu_m$ is a positive measure on $\Omega$, whereas (iii) implies
that there is no mass present in a set that has no volume (w.r.t.
$\lambda^d$). A mass measure satisfying Postulate \ref{Postulate
mu_m} is in this context referred to as a \textit{macroscopic mass
measure}. Radon-Nikodym Theorem\footnote{See \cite{Bradley} for a variant of this Theorem
formulated for finite measures which is applied here.} (see \cite{Halmos} for more details on this subject)
guarantees the existence of a real, non-negative density
$\hat{\rho}\in L_{\lambda^d}^1(\Omega)$ such that:
\begin{equation}
\mu_m(\Omega')=\int_{\Omega'} \hat{\rho}(x)d\lambda^d(x) \hspace{1
cm} \text{for all } \Omega'\in\mathcal{B}(\Omega).
\end{equation}

Similarly, we introduce time-dependent mass measures $\mu_t$, where
the time slice $t\in S$ enters as a parameter.

\subsection{Porosity measure}\label{section Porosity measure}

Let $\Omega\subset\mathbb{R}^d$ be a heterogeneous domain composed
of two distinct regions: {\em free space} for pedestrian motion and a {\em matrix} (obstacles) such
that $\Omega=\Omega_s\cup\Omega_p$ (disjoint union), where
$\Omega_s$ is the matrix (solid part) of $\Omega$ and $\Omega_p$ is
the free space (pores). This notation is very much inspired by the
modeling of transport and chemical reactions in porous media; see
\cite{Bear}, e.g.

Let $\mu_p(\Omega')$ be the volume of pores in $\Omega'\subset\Omega$.
\begin{postulate}[Assumptions on $\mu_p$]\label{Postulate mu_p}
\begin{enumerate}
  \item[(i)] $\mu_p\geqslant 0$.\label{Postulate mu_p 1}\\
  \item[(ii)] $\mu_p$ is $\sigma$-additive.\label{Postulate mu_p 2}\\
  \item[(iii)] $\mu_p\ll\lambda^d$.\label{Postulate mu_p 3}
\end{enumerate}
\end{postulate}
By  Postulate \ref{Postulate mu_p} (i) and (ii), we have that
$\mu_p$ is a measure on $\Omega$. We  refer to $\mu_p$ as a
\textit{porosity measure} (cf. \cite{Boehm}). The absolute
continuity statement in (iii) formulates  mathematically that there
cannot be a non-zero volume of pores included in a set that has zero
volume (w.r.t. $\lambda^d$). Assume that $\Omega$ is such that
$\lambda^d(\Omega)<\infty$. Then the Radon-Nikodym Theorem ensures
the existence of a function $\phi\in L^1_+(\Omega)$ such that:
\begin{equation}
\mu_p(\Omega')=\int_{\Omega'}\phi d\lambda^d \quad \mbox{ for all }
\Omega'\in\mathcal{B}(\Omega).
\end{equation}
 Note that $\mu_p(\Omega')$
measures the volume of a subset of $\Omega'$ (namely of
$\Omega'\cap\Omega_p$). So, we get that
\begin{equation}
\mu_p(\Omega')=\lambda^d(\Omega'\cap\Omega_p)\leqslant\lambda^d(\Omega')
\quad \mbox{ for all } \Omega'\in\mathcal{B}(\Omega).
 \end{equation}
We thus have $\int_{\Omega'}\phi d\lambda^d\leqslant \int_{\Omega'}d\lambda^d$, or $\int_{\Omega'}(1-\phi)d\lambda^d\geqslant0$. Since the latter inequality holds for any choice of $\Omega'$, it follows that $\phi\leqslant 1$ almost everywhere in $\Omega$.

\subsection{Transport of a measure}

For the sequel, we wish to restrict the presentation to the case $d=2$. For our time interval $S$ and for each $i\in\{1,2\}$, we  denote the velocity field of the corresponding measure by $v^i(t,x)$ with $(t,x)\in S\times\Omega$. Let also $\mu_t^1$, and $\mu_t^2$ be two time-dependent mass measures. Note that for each choice of $i$, the dependence on $t$ of $v^i$ is comprised in  the functional dependence of $v^i$  on both measures $\mu_t^1$, and $\mu_t^2$. This is clearly indicated in (\ref{Def v decomposition}).
The fact that here we deal with two mass measures $\mu_t^1$ and $\mu_t^2$, transported with corresponding velocities $v^1$ and $v^2$, translates into:
\begin{equation}\label{Cons of mass}
 \left\{
    \begin{aligned}
        \dfrac{\partial \mu_t^1}{\partial t}+\nabla \cdot (\mu_t^1 v^1)&=0,\\
        \dfrac{\partial \mu_t^2}{\partial t}+\nabla \cdot (\mu_t^2 v^2)&=0,
    \end{aligned}
 \right. \hspace{1 cm}\text{for all }(t,x)\in S \times\Omega.
\end{equation}
These equations are accompanied by the following set of initial conditions:
\begin{equation}
\mu_t^i=\mu_0^i \mbox{ as } t=0 \mbox{ for } i\in\{1,2\}.
\end{equation}
It is worth noting that (\ref{Cons of mass}) is the measure-theoretical counterpart of the Reynolds Theorem in continuum mechanics.
To be able to interpret what a partial differential equation in
terms of measures means, we give a weak formulation of (\ref{Cons of
mass}).
Essentially, for all test functions $\psi^1, \psi^2\in
C_0^1(\bar{\Omega})$  and for almost every $t\in S$, the following
identity holds:
\begin{equation}\label{Weak Form}
\dfrac{d}{dt}\int_{\Omega}\psi^i(x)d\mu_t^i(x)=\int_{\Omega}v^i(t,x)\cdot\nabla\psi^i(x)d\mu_t^i(x)\mbox{
for all } i\in\{1,2\}.
\end{equation}

\begin{definition}[Weak solution of (\ref{Cons of mass})]
The pair $(\{\mu_t^1\}_{t\geqslant 0},\{\mu_t^2\}_{t\geqslant 0})$
is called a weak solution of (\ref{Cons of mass}), if for all
$i\in\{1,2\}$ the following properties hold:
\begin{enumerate}
\item the mappings $t\mapsto \int_{\Omega}\psi^i(x)d\mu_t^i(x)$ are absolutely continuous for all $\psi^i\in C_0^1(\bar{\Omega})$;
\item $v^i\in L^2\Bigl(S;L_{\mu_t^i}^1(\Omega)\Bigr)$;
\item Equation (\ref{Weak Form}) is fulfilled.
\end{enumerate}
\end{definition}
We refer the reader to \cite{Canuto} for an example where the
existence of weak solutions to a similar (but easier) transport equation for
measures has been rigorously shown.

\section{Social velocities}\label{social}

We follow very much the philosophy  developed by Helbing, Vicsek and
coauthors (see, e.g. \cite{HelbingVicsek} and references cited
therein) which defends the idea that the pedestrian's motion is
driven by a social force. Is worth noting that similar thoughts were
given in this direction (motion of social masses/networks) much
earlier, for instance,   by Spiru Haret \cite{Haret} and Antonio
Portuondo y Barcel\'o \cite{Apuntes}. Moreover, other authors (for instance, Hoogendoorn and Bovy \cite{optimal}) prefer to account also for the Zipfian principle of least effort for the human behavior. We do not attempt to capture the least effort principle in this study.

\subsection{Specification of the velocity fields $v^i$}\label{section Specify velocity}
Until now, we have not explicitly defined the velocity fields $v^i$ ($i\in\{1,2\}$). Very much inspired by the \textit{social force model} by Dirk Helbing \textit{et al.} \cite{HelbingMolnar}, the velocity of a pedestrian is modeled as a \textit{desired velocity} $v_{\text{des}}^i$ perturbed by a component $v_{[\mu_t^1,\mu_t^2]}^i$. The latter component is due to the presence of other individuals, both from the pedestrian's own subpopulation and from the other subpopulation. The desired velocity is independent of the measures $\mu_t^1$ and $\mu_t^2$, and represents the velocity that an agent would have had in absence of other pedestrians.

For each $i\in\{1,2\}$,  the velocity $v^i$ is   defined by superposing
the two velocities $v_{\text{des}}^i$ and $v_{[\mu_t^1,\mu_t^2]}^i$ as follows:
\begin{equation}\label{Def v decomposition}
v^i(t,x):= v_{\text{des}}^i(x)+v_{[\mu_t^1,\mu_t^2]}^i(x), \hspace{1
cm} \text{for all }t\in(0,T) \text{ and }x\in\Omega.
\end{equation}
For a \textit{counterflow} scenario, the desired velocities of the
two subpopulations follow opposite directions. We thus take
$$v_{\text{des}}^i(x)=v_{\text{des}}^i\in\mathbb{R}^2$$ fixed (for
$i\in\{1,2\}$) and $$v_{\text{des}}^1=-v_{\text{des}}^2.$$ The
component $v_{[\mu_t^1,\mu_t^2]}^i$ models the effect of
interactions with other pedestrians on the current
velocity\footnote{The interactions we are pointing at are {\em
nonlocal}.}. Since the interactions between members of the same
subpopulation differ (in general) from the interactions between
members of opposite subpopulations, we assume that
$v_{[\mu_t^1,\mu_t^2]}^i$ consists of two parts:
\begin{eqnarray}\label{Def v measure}
\nonumber v_{[\mu_t^1,\mu_t^2]}^i(x)&:=&\int_{\Omega \setminus \{x\}} f^{\text{own}}(|y-x|)g(\alpha_{xy}^i)\dfrac{y-x}{|y-x|}d\mu_t^i(y)\\
&& + \int_{\Omega \setminus \{x\}}
f^{\text{opp}}(|y-x|)g(\alpha_{xy}^i)\dfrac{y-x}{|y-x|}d\mu_t^j(y),
\end{eqnarray}
for $i\in\{1,2\}$, where $j=1$ if $i=2$ and vice versa.
In (\ref{Def v measure}) we have used the following:
\begin{itemize}
  \item $f^{\text{own}}$ and $f^{\text{opp}}$ are continuous functions from $\mathbb{R}_+$ to $\mathbb{R}$,
  describing the effect of the mutual distance between individuals on their interaction. Compare the concept of {\em distance interactions} defined in \cite{interactions}.
   $f^{\text{own}}$ incorporates the influence by members of the same subpopulation, whereas $f^{\text{opp}}$ accounts for the interaction between members of opposite subpopulations. $f^{\text{own}}$ is a composition of two effects: on the one hand individuals are repelled, since they want to avoid collisions and congestion, on the other hand they are attracted to other group mates, in order not to get separated from the group. $f^{\text{opp}}$ only contains a repulsive part, since we assume that pedestrians do not want to stick to the other subpopulation.
  \item $\alpha_{xy}^i$ denotes the angle between $y-x$ and $v_{\text{des}}^i(x)$: the angle under which $x$ sees $y$ if it were moving in the direction of $v_{\text{des}}^i(x)$.
  \item $g$ is a function from $[-\pi,\pi]$ to $[0,1]$ that encodes the fact that an individual's vision is not equal in all directions.
\end{itemize}

 Regarding the specific choice of $f^{\text{own}}$, $f^{\text{opp}}$ and $g$ we are very much inspired by \cite{HelbingMolnar} and \cite{Piccoli2010}, e.g. However we do not use exactly their way of modeling pedestrians' interaction forces. We list here the following forms for the functions $f^{\text{own}}$, $f^{\text{opp}}$ and $g$ that match the given characterization:
\begin{eqnarray}
f^{\text{opp}}(s)&:=&\left\{
  \begin{array}{ll}
    -F^{\text{opp}} \Bigl(\dfrac{1}{s^2}-\dfrac{1}{(R_r^{\text{opp}})^2}\Bigr), & \mbox{if $s\leqslant R_r^{\text{opp}}$;} \\
    0, & \mbox{if $s>R_r^{\text{opp}}$,}
  \end{array}
\right.\label{fopp explicit}\\
f^{\text{own}}(s)&:=&\left\{
  \begin{array}{ll}
    -F^{\text{own}} \Bigl(\dfrac{1}{s}-\dfrac{1}{R_r^{\text{own}}}\Bigr)\Bigl(\dfrac{1}{s}-\dfrac{1}{R_a^{\text{own}}}\Bigr), & \mbox{if $s\leqslant R_a^{\text{own}}$;} \\
    0, & \mbox{if $s>R_a^{\text{own}}$,}
  \end{array}
\right.\label{fown explicit}\\
g(\alpha)&:=& \sigma +(1-\sigma)\dfrac{1+\cos(\alpha)}{2}, \hspace{1 cm}\text{for }\alpha\in[-\pi.\pi].\label{g(alpha)}
\end{eqnarray}
Here $F^{\text{opp}}$ and $F^{\text{own}}$ are fixed, positive constants. The constants $R_r^{\text{opp}}$, $R_r^{\text{own}}$ (\textit{radii of repulsion}) and $R_a^{\text{own}}$ (\textit{radius of attraction}) are fixed and should be chosen such that $0<R_r^{\text{own}}<R_a^{\text{own}}$ and $0<R_r^{\text{opp}}$. Furthermore,  the restriction $\max\{R_a^{\text{own}}, R_r^{\text{opp}}\}\ll L$ has to be fulfilled. The interaction $f^{\text{opp}}$ is designed such that individuals ``feel" repulsion (i.e. $f^{\text{opp}}<0$) from another pedestrian if they are placed within a distance $R_r^{\text{opp}}$ from one another. The corresponding statement  holds for $f^{\text{own}}$ if individuals are within the distance $R_r^{\text{own}}$. Additionally,  an individual is attracted to a second individual if their mutual distance ranges between $R_r^{\text{own}}$ and $R_a^{\text{own}}$.\\
The function $g$ ensures that an individual experiences the strongest influence from someone straight ahead, since $g(0)=1$ for any $\sigma\in[0,1]$. The constant $\sigma$ is a tuning parameter  called \textit{potential of anisotropism}. It determines how strongly a pedestrian is focussed on what happens in front of him, and how large the influence is of people at his sides or behind him.

In the remainder of this section, we suggest four different
alternatives for the definition of $v_{[\mu_t^1,\mu_t^2]}^i$ by
indicating various special choices of distance interactions  and visibility angles (conceptually similar to
$\alpha_{xy}^i$) as they arise in (\ref{Def v measure}).
 All of them boil down to including an implicit dependency of the actual velocity $v^i= v_{\text{des}}^i+v_{[\mu_t^1,\mu_t^2]}^i$.
Note that this effect increases the degree of realism of the model,
but on the other hand it makes the mathematical justification of the
corresponding models much harder to get.

\subsubsection{Modification of the angle $\alpha_{xy}^i$}

We defined the angle  $\alpha_{xy}^i$ as the angle between the
vector $y-x$ and $v_{\text{des}}^i(x)$. However this is not a good
definition if the pedestrian in position $x$ is not moving in the
direction of $v_{\text{des}}^i(x)$ (or, in a broader sense, if the
actual speed cannot be approximated sufficiently well by the desired
velocity). Therefore we suggest to define
$\alpha_{xy}^i=\alpha_{xy}^i(t)$ as the angle between $y-x$ and
$v^i(t,x)$.

\subsubsection{Prediction of mutual distance in (near) future}

Up to now the functions $f^{\text{own}}$ and $f^{\text{opp}}$
depended on the actual distance between $x$ and $y$ at time $t$.
However pedestrians are likely to anticipate on the distance they
expect to have after a certain (small) time-step (say, some fixed
$\Delta t \in \mathbb{R}$). In practice, this means that at a time
$t\in S$ a person will modify his velocity (either in direction, or
in magnitude, or both) if he foresees a collision at time $t+\Delta
t\in S$.

To predict the mutual distance between $x$ and $y$ at time $t+\Delta
t$, the current velocities at $x$ and $y$ are used for
extrapolation. The predicted distance is: $|(y+v(y,t)\Delta
t)-(x+v(x,t)\Delta t)|$. Consequently, sticking to the notation in
(\ref{Def v measure}), the interaction potential $f^{\text{own}}$
and $f^{\text{opp}}$ should depend on $|(y+v^i(t,y)\Delta
t)-(x+v^i(t,x)\Delta t)|$ and on $|(y+v^j(t,y)\Delta
t)-(x+v^i(t,x)\Delta t)|$ respectively (where $j=1$ if $i=2$ and vice versa).

\subsubsection{Prediction of mutual distance within a time interval
in the (near) future}

The disadvantage of using $|(y+v(y,t)\Delta t)-(x+v(x,t)\Delta t)|$
is that $\Delta t$ is fixed. A pedestrian can thus only predict the
distance at an \textit{a priori} specified point in time in the
future. However, people are able to anticipate also if they expect a
collision to occur at a time that is not equal to $t+\Delta t$. We
assume now that we are given a fixed $\Delta
t_{\text{max}}\in\mathbb{R}_+$ such that an individual can predict
mutual distances by extrapolation for any time $\tau\in(t, t+\Delta
t_{\text{max}})$. Thus, $\Delta t_{\text{max}}$ imposes a bound on
how far can an individual look ahead into the future.
To capture this effect,
we suggest to replace $f^{\text{own}}(|y-x|)$  and $f^{\text{opp}}(|y-x|)$ by:
 \begin{equation}\label{11}
\frac{1}{\Delta
t_{\text{max}}}\int_0^{\Delta
t_{\text{max}}}f^{\text{own}}(\big|(y+v^i(t,y)\tau)-(x+v^i(t,x)\tau)\big|)d\tau,
\end{equation}
and
\begin{equation}\label{12}
\frac{1}{\Delta
t_{\text{max}}}\int_0^{\Delta
t_{\text{max}}}f^{\text{opp}}(\big|(y+v^j(t,y)\tau)-(x+v^i(t,x)\tau)\big|)d\tau,
\end{equation}
respectively.

\subsubsection{Weighted prediction}

Since an individual probably attaches more value to his predictions for points in time that are nearer by than others, one additional modification comes to our mind.
 Let $h:[t,t+\Delta t_{\text{max}}]\rightarrow[0,1]$ be a weight
 function. Then instead of (\ref{11}) and (\ref{12}), we propose
\begin{equation}
\frac{1}{\Delta
t_{\text{max}}}\int_0^{\Delta
t_{\text{max}}}f^{\text{own}}(\big|(y+v^i(t,y)\tau)-(x+v^i(t,x)\tau)\big|)h(\tau)d\tau,
\end{equation}
and
\begin{equation}
\frac{1}{\Delta
t_{\text{max}}}\int_0^{\Delta
t_{\text{max}}}f^{\text{opp}}(\big|(y+v^j(t,y)\tau)-(x+v^i(t,x)\tau)\big|)h(\tau)d\tau.
\end{equation}
If $h$ is decreasing, then the influence of $t_1$ is larger than the
influence of $t_2$, if $t_1<t_2$ (which matches our intuition).

\subsection{Two-scale measures}

We now consider the explicit decomposition of the measures $\mu_t^1$
and $\mu_t^2$. Let the pair $(\theta_1,\theta_2)$ be in $[0,1]^2$,
and consider the following decomposition of $\mu_t^1$ and $\mu_t^2$:
\begin{equation}\label{decomposition mass measure micro macro}
\mu_t^i=\theta_i m_t^i + (1-\theta_i)M_t^i, \hspace{1 cm} i\in
\{1,2\}.
\end{equation}
Here, $m_t^i$ is a microscopic measure. We consider
$\{p_k^i(t)\}_{k=1}^{N^i}\subset\Omega$ to be the positions at time
$t$ of $N^i$ chosen pedestrians, that are members of subpopulation
$i$. We want $m_t^i$ to be a counting measure with respect to these
pedestrians, i.e. for all $\Omega'\in\mathcal{B}(\Omega)$:
\begin{equation}
m_t^i(\Omega')=\#\{p_k^i(t)\in \Omega'\}, \hspace{1 cm} i\in\{1,2\}.
\end{equation}
We thus define $m_t^i$ as the sum of Dirac masses (cf. Section
\ref{section micro mass measure}), centered at the $p_k^i$,
$k=1,2,\ldots,N^i$:
\begin{equation}
m_t^i=\sum_{k=1}^{N^i} \delta_{p_k^i(t)}, \hspace{1 cm} i\in\{1,2\}.
\end{equation}
$M_t^i$ is the macroscopic part of the measure, which takes into
account the part of the crowd that is considered continuous. We
consequently have $M_t^i\ll \lambda^2$, since a set of zero volume
cannot contain any mass. Note that we are thus in the setting of
Section \ref{section macro mass measure}. Now, Radon-Nikodym
Theorem guarantees the existence of a real,
non-negative density $\hat{\rho}^i(t,\cdot)\in
L_{\lambda^2}^1(\Omega)$ such that:
\begin{equation}
M_t^i(\Omega')=\int_{\Omega'} \hat{\rho}^i(t,x)d\lambda^2(x) \end{equation}
for all  $\Omega'\in\mathcal{B}(\Omega)$ and
 all  $i\in\{1,2\}$.

\section{Micro-macro modeling of pedestrians motion in heterogeneous
domains}\label{micro-macro}

We have already made clear that we want to model the heterogeneity
of the interior of the corridor. In practice this means that
pedestrians cannot enter all parts of the domain. As described in
Section \ref{section Porosity measure}, we have a measure $\mu_p$
corresponding to the porosity of the domain (which is fixed in
time). However, we note that the concept of porosity (cf. Section
\ref{section Porosity measure}) is a macroscopic one. For this
reason only the macroscopic part of the mass measure in
(\ref{decomposition mass measure micro macro}) needs some
modification with respect to the porosity. In this context, one
should be aware of the analogy with mathematical
\textit{homogenization}. This technique distinguishes between
microscopic and macroscopic scales, where we also see that some
(averaged) characteristics are only defined on the macroscopic
scale. For more details, the reader is referred to \cite{Bear} or
\cite{Jikov}. In $\mathbb{R}^2$, we have $\mu_p\ll \lambda^2$.
Furthermore $M_t^i\ll \mu_p$ for  $i\in \{1,2\}$ and a.e. $t\in S$.
This is obvious, since no pedestrians can be present in a set that
has no pore space (i.e. zero porosity measure). A basic property of Radon-Nikodym derivatives now gives us:
\begin{equation}
\dfrac{d
M_t^i}{d\lambda^2}=\dfrac{dM_t^i}{d\mu_p}\dfrac{d\mu_p}{d\lambda^2}
\quad i\in\{1,2\} \text{ for almost every } t\in S.
\end{equation}
We have already defined $\hat{\rho}^i(t,\cdot):=\dfrac{d
M_t^i}{d\lambda^2}$ and $\phi:=\dfrac{d\mu_p}{d\lambda^2}$. If we
now denote by $\rho^i(t,\cdot)$ the Radon-Nikodym derivative
$\dfrac{dM_t^i}{d\mu_p}$, the following relation holds:
$\hat{\rho}^i(t,\cdot)\equiv\rho^i(t,\cdot)\phi(\cdot)$ for all
$i\in\{1,2\}$.

\subsection{Weak formulation for micro-macro mass measures}

We now have the following measure:
\begin{equation}
\mu_t^i=\theta_i m_t^i + (1-\theta_i)M_t^i, \hspace{1 cm} i\in
\{1,2\},
\end{equation}
as was given in (\ref{decomposition mass measure micro macro}),
where now:
\begin{equation}
m_t^i=\sum_{k=1}^{N^i} \delta_{p_k^i(t)}, \hspace{1 cm}
dM_t^i(x)=\rho^i(t,x)\phi(x)d\lambda^2(x).
\end{equation}
This specific form of the measure will now be included in the weak
formulation (\ref{Weak Form}), with velocity field (\ref{Def v
decomposition})-(\ref{Def v measure}). The real positive numbers $\theta_i$ ($i\in\{1,2\}$) are intrinsic scaling parameters depending on $N^i$.

The transport equation
(\ref{Weak Form}) takes the following form:

\begin{align}
\nonumber &\dfrac{d}{dt}\Bigl(\theta_i \sum_{k=1}^{N^i} \psi^i\bigl(p_k^i(t)\bigr) + (1-\theta_i) \int_{\Omega}\psi^i(x)\rho^i(t,x)\phi(x)d\lambda^2(x)\Bigr)=\\
& \theta_i \sum_{k=1}^{N^i} v^i\bigl(t,p_k^i(t)\bigr)\cdot\nabla
\psi^i\bigl(p_k^i(t)\bigr)+(1-\theta_i)
\int_{\Omega}v^i(t,x)\cdot\nabla\psi^i(x)\rho^i(t,x)\phi(x)d\lambda^2(x),
\end{align}
for all  $i\in\{1,2\}$. Here we have used the sifting property of
the Dirac distribution. In the same manner, we specify
$v_{[\mu_t^1,\mu_t^2]}^i$ from (\ref{Def v measure}) as
\begin{eqnarray}
\nonumber v_{[\mu_t^1,\mu_t^2]}^i(x)&=& \theta_i\sum_{\substack{k=1\\p_k^i(t)\neq x}}^{N^i}f^{\text{own}}(|p_k^i(t)-x|)g(\alpha_{xp_k^i(t)}^i)\dfrac{p_k^i(t)-x}{|p_k^i(t)-x|}\\
\nonumber
&& + (1-\theta_i)\int_{\Omega} f^{\text{own}}(|y-x|)g(\alpha_{xy}^i)\dfrac{y-x}{|y-x|}\rho^i(t,y)\phi(y)d\lambda^2(y)\\
\nonumber && +\theta_j\sum_{\substack{k=1\\p_k^j(t)\neq x}}^{N^j}f^{\text{opp}}(|p_k^j(t)-x|)g(\alpha_{xp_k^j(t)}^i)\dfrac{p_k^j(t)-x}{|p_k^j(t)-x|}\\
&& + (1-\theta_j)\int_{\Omega}
f^{\text{opp}}(|y-x|)g(\alpha_{xy}^i)\dfrac{y-x}{|y-x|}\rho^j(t,y)\phi(y)d\lambda^2(y),\nonumber
\end{eqnarray}
for $i\in \{1,2\}$, and $j$ as before ($j=1$ if $i=2$ and vice
versa). We have omitted the exclusion of $\{x\}$ from the domain of
integration (in the macroscopic part), since $\{x\}$ is a nullset
and thus negligible w.r.t. $\lambda^2$. Note that the sums may be evaluated in any
point $x\in\Omega$ (not necessarily $x=p_k^i(t)$ for some $i$ and
$k$); the integral parts may also be evaluated in all $x$, including
$x=p_k^i(t)$ for some $i$ and $k$.

\section{Numerical illustration}\label{simulation}

 We wish to illustrate now the microscale description of a counterflow scenario (i.e. for $\theta_1=\theta_2=1$) by presenting plots of the configuration of all individuals situated in a given corridor at specific moments in time.

We consider a specific instance in which there are in total 40 individuals (20 in each subpopulation). The dimensions of the corridor are $d=4$ and $L=20$. The velocity is taken as defined in (\ref{Def v measure})-(\ref{g(alpha)}). Furthermore, the following model parameters are used:
$v_{\text{des}}^1 = 1.34e_1$,
$v_{\text{des}}^2 = -1.34e_1$, $F^{\text{opp}} = 0.3$, $F^{\text{own}} = 0.3,$ $R_r^{\text{opp}} = 2,$
$R_a^{\text{own}} = 2,$
$R_r^{\text{own}} = 0.5$, $F^{\text{w}} = 0.5$, $R^{\text{w}} = 0.5,$
$\sigma = 0.5.$

\begin{figure}[t]
\vspace{-2 cm}
\begin{tabular}{lcr}
\hspace{-1.00 cm}\includegraphics[width=0.4\linewidth]{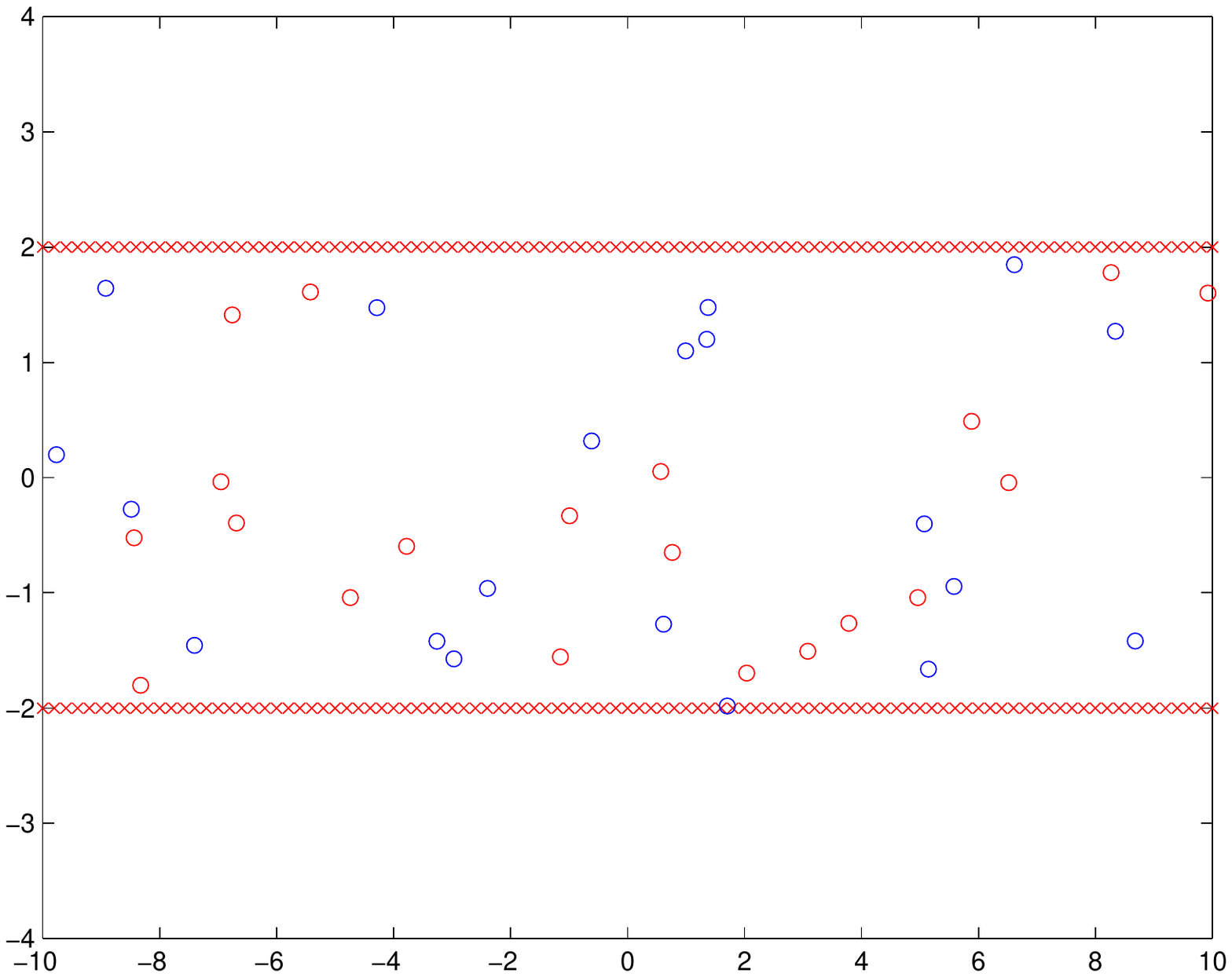}
&
\hspace{-0.85 cm}\includegraphics[width=0.4\linewidth]{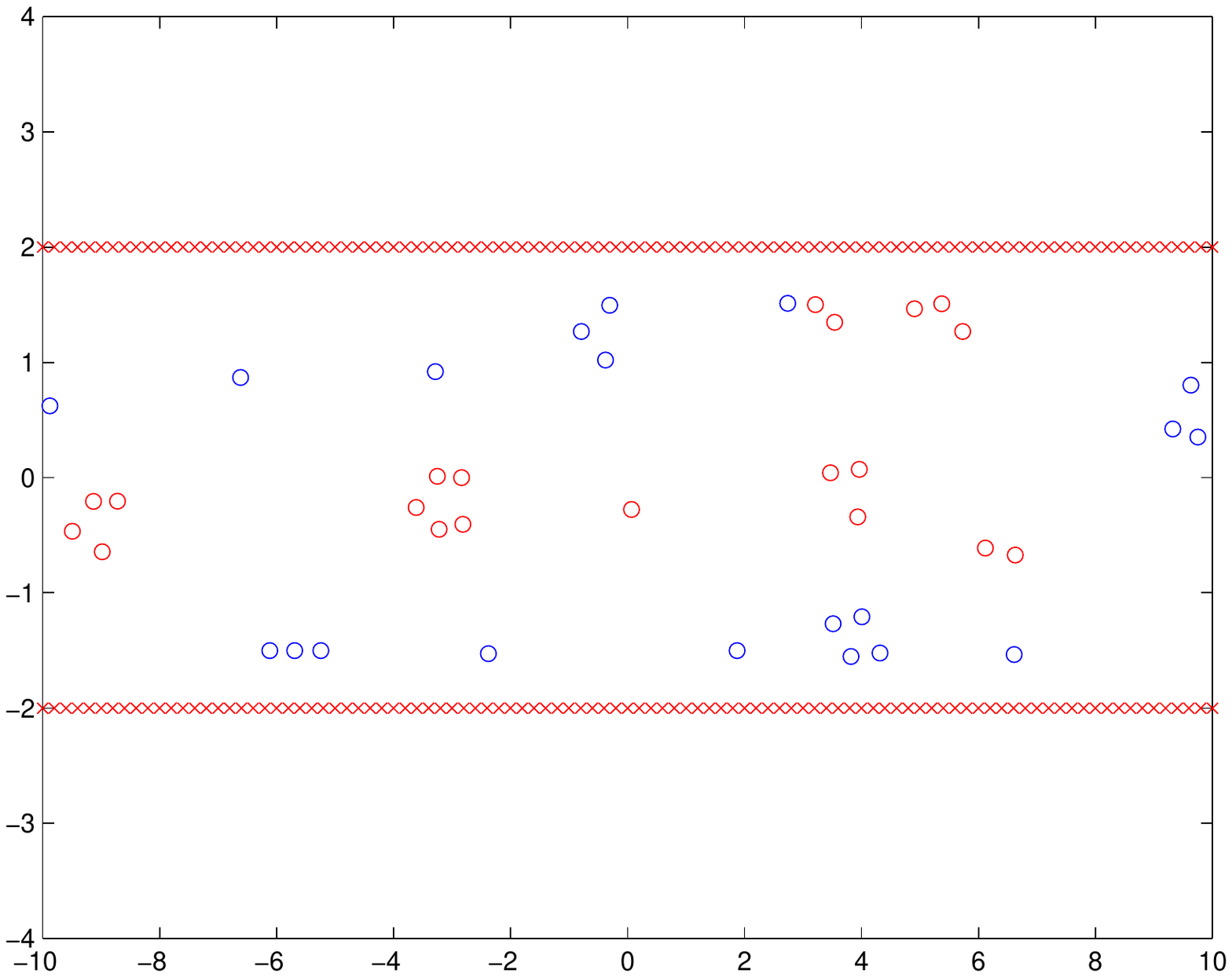}
&
\hspace{-1.05 cm}\includegraphics[width=0.4\linewidth]{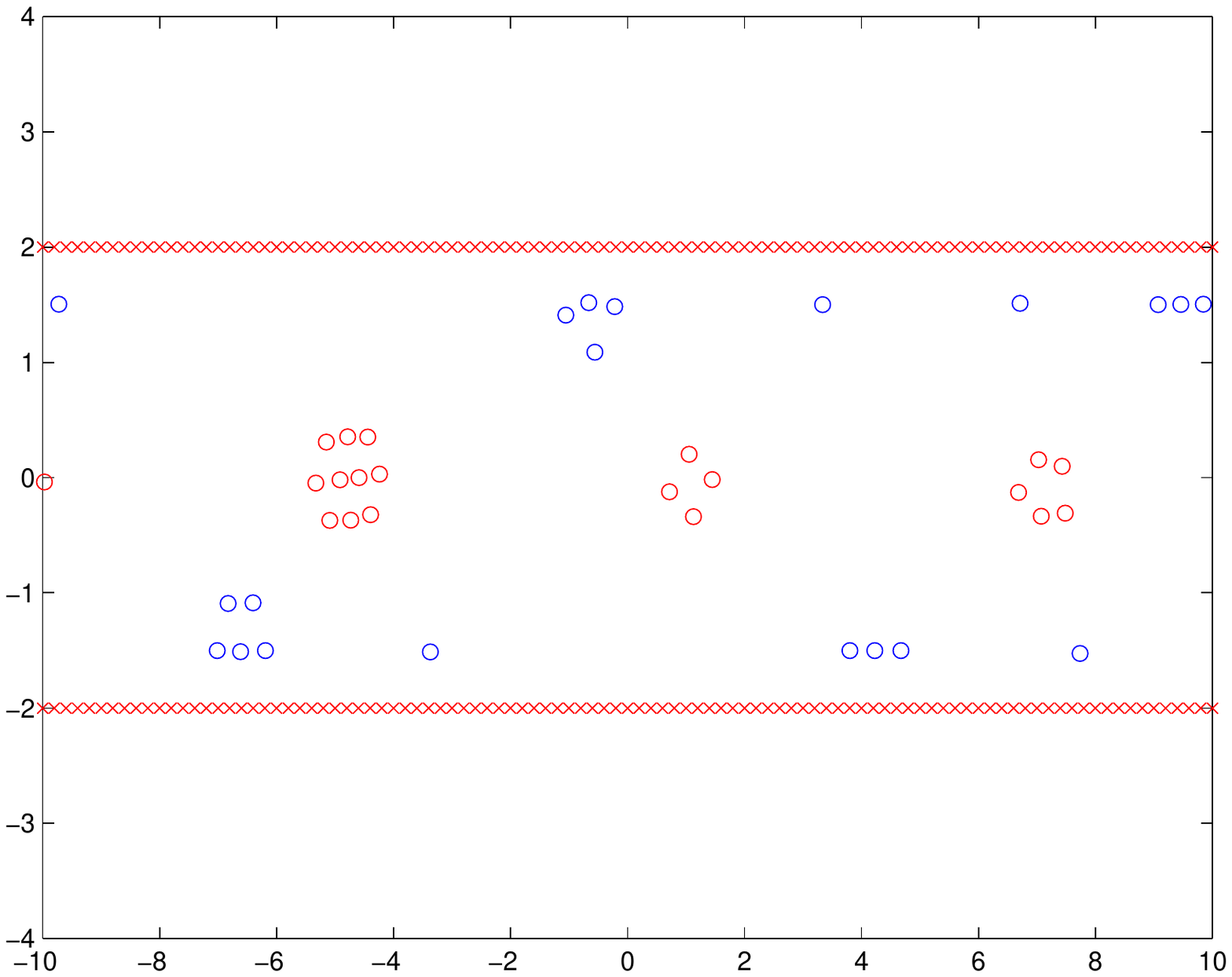} \\
\end{tabular}
\vspace{-2 cm}
\caption{The simulation of a crowd's motion in a corridor of length $L=20$ and width $d=4$. Each of the two sub-populations consists of 20 individuals. The images were taken at $t=0$ (left), $t=7.5$ (middle), $t=15$ (right).}\label{graph typical example}
\end{figure}

In Figure \ref{graph typical example}, we show the configuration in the corridor at times $t=0$, $t=7.5$, and $t=15$. The individuals of the subpopulation 1 are colored blue, while the individuals of the subpopulation 2 are colored red. Clearly, self-organization can be observed in the system: Pedestrians that desire to move in the same direction form lanes (in this case, three of them). This feature is observed and described extensively in literature, cf. e.g. \cite{HelbingVicsek}.

Another feature, pointed out by Figure \ref{graph typical example}, is the following: Within the three already formed  lanes, small clusters of people are formed. This flocking is a result of the typical choice for $f^{\text{own}}$ in (\ref{fown explicit}). Members of the same subpopulation are repelled if their mutual distance is in the range $(0,R_r^{\text{own}})$; they are attracted if their mutual distance is in the range $(R_r^{\text{own}},R_a^{\text{own}})$. No interaction takes place if individuals are more than a distance $R_a^{\text{own}}$ apart.

The attraction part of the interaction causes individuals that are already relatively close to get even closer, until they are at a distance $R_r^{\text{own}}$. For distances around $R_r^{\text{own}}$, there is an interplay between repulsion and attraction, eventually leading to some equilibrium in the mutual distances between neighboring individuals in one cluster. In Figure \ref{graph typical example},  we observe self-organized patterns even within clusters.

\section*{Acknowledgments}
We acknowledge fruitful discussions within the "Particle Systems
Seminar" of ICMS (Institute for Complex Molecular Systems, TU
Eindhoven, The Netherlands), especially with H. ten Eikelder, B.
Markvoort, F. Nardi, M. Peletier, M. Renger, and F. Toschi. A.M. is indebted to
Michael B\"ohm (Bremen) for introducing him to the fascinating world
of modeling with measures.

\bibliographystyle{abbrv}
\bibliography{refs}

\end{document}